\documentclass[11pt]{article}
\usepackage{moriond,epsfig}

\def\be{\begin{equation}}
\def\ee{\end{equation}}
\def\bea{\begin{eqnarray}}
\def\eea{\end{eqnarray}}
\def\lsim{\mathrel{\rlap{\lower4pt\hbox{\hskip1pt$\sim$}}
    \raise1pt\hbox{$<$}}} 
\def\gsim{\mathrel{\rlap{\lower4pt\hbox{\hskip1pt$\sim$}}
    \raise1pt\hbox{$>$}}}  

\begin{document}
\vspace*{4cm}
\title{Time Delay Interferometry and LISA Optimal Sensitivity}
\author{A.~Pai\dag, K.~Rajesh~Nayak\ddag, S.~V.~Dhurandhar\ddag~ and J-Y.~Vinet\dag}
\address{\dag CNRS, Observatoire de la C\^ote d'Azur
UMR6162--ILGA (Interf\'erom\'etrie Laser pour la Gravitation et l'Astrophysique)
BP 4229 F-06304 Nice Cedex 4 France.\\
\ddag Inter-University Centre for Astronomy and Astrophysics, Pune, 
India.}

\maketitle\abstract{
The sensitivity of LISA depends
on the suppression of several noise sources; dominant one is laser
frequency noise. It has been shown that the six Doppler data streams
obtained from three space-crafts can be appropriately time delayed and
optimally combined to cancel this laser frequency noise. We show that the optimal
data combinations when operated in a {\it network} mode improves the
sensitivity over Michelson ranging from $40 \%$ to $100 \%$.
In this article, we summarize these results. We further show that the
residual laser noise in the optimal data combination due to typical arm-length inaccuracy of $200$ m
is much below the level of optical path and the proof mass noises.} 

\section{Introduction}
The future space-based gravitational wave (GW) mission \cite{LISArep}- the 
Laser Interferometric Space Antenna (LISA) - consists of three identical
space-crafts forming an equilateral triangle of side $5\times 10^{6}$ km 
following heliocentric orbits trailing the Earth by $20^\circ$. LISA is
thus a giant interferometric configuration with
three arms which will give independent information on GW polarizations
and will detect GW in the low frequency range of 0.1 mHz to 0.1 Hz.
Due to the long arm-lengths of the antenna, it is not feasible 
to bounce the laser beams. A special
Doppler tracking scheme is used to track the space-crafts with laser
beams. This exchange of laser beams between the three space-crafts 
result in six Doppler data streams.
The LISA sensitivity is limited by many noise sources; the dominant
one is the laser phase noise; noise due to phase fluctuations of the
master laser. The current stabilization schemes estimate this
noise to about $\Delta \nu /\nu_0 \simeq 10^{-13}/\sqrt{{Hz}}$, where $\nu_0$ is the frequency of the 
laser and $\Delta \nu$ the fluctuation in frequency.
If the laser frequency noise can be suppressed then the noise floor is determined by the  
optical-path noise which fakes the fluctuations in the lengths of optical
paths and the residual acceleration of proof masses resulting from imperfect 
shielding of the drag-free system. 
Thus, canceling the laser frequency noise is vital for LISA to
reach the requisite sensitivity of $h \sim 10^{-21}$ or $10^{-22}$. Since it is impossible to maintain equal distances between space-crafts, 
cancellation of laser frequency noise is a non-trivial problem.  
Several schemes have been proposed to combat this noise. 
In these schemes \cite{AET99,ETA00}, 
the data streams are combined with appropriate time delays in order to
cancel the laser frequency noise. In our earlier work, it was
rigorously shown that {\it all} laser-noise free data combinations form an
algebraic module over a polynomial ring over time delay operators \cite{SNV02}. Furthermore,
recently \cite{SNV03}, we have found data combinations which are eigen-data combinations of the noise covariance matrix, which give optimal sensitivity, when averaged
over all the source directions and polarization angles. We show that
signal-to-noise ratio (SNR) for any data
combination in the laser noise-free cancellation module lies between
the SNR's of these eigen-data combinations. In this
article, we summarize our previous results
\cite{SNV02,SNV03}. Besides, we estimate the residual laser noise due
to inaccuracies in the arm-lengths.  

\section{Time Delay Interferometry and Laser Noise Cancellation Module}\label{}
\begin{figure}
\begin{center}
\includegraphics[width=6cm,height=6cm]{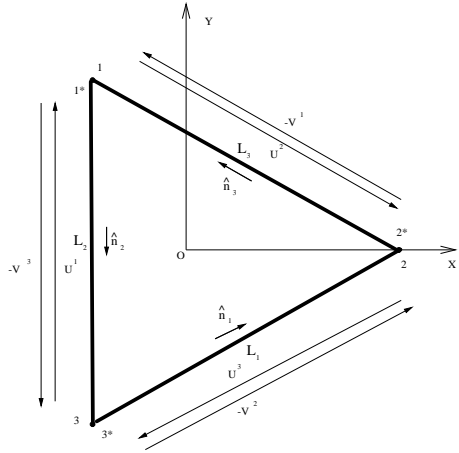}
\caption{The LISA constellation}
\label{fig:lisa}
\end{center}
\end{figure}

In LISA constellation, the six Doppler data streams are labeled as
$U^i$ and $V^i, i = 1, 2, 3$; if the space-crafts are labeled
clockwise as shown in Fig. \ref{fig:lisa}. The Doppler data stream $U^1$ is obtained by
letting the laser beam from space-craft 3 to travel towards space-craft 1 along the arm of length $L_2$ in the 
direction $- \hat{\bf n}_2$, and is beaten with the on-board laser beam
at space-craft 1. Similarly, $-V^1$ represents the Doppler data
obtained by beating laser beam traveling from space-craft 2 to
space-craft 1 along the arm of length $L_3$ in the direction of
$\hat{\bf n}_3$ with the
on-board laser at space-craft 1. The remaining 4 beams are described by cyclically 
permuting the indices. These beams contain the laser frequency noise, other noises such as optical path, acceleration etc. 
and also the GW signal. 
Using time delay interferometry, these $6$
Doppler data streams can be appropriately combined by delaying them
with time delay operators
$\,E_i x (t) = x (t - L_i), i = 1, 2, 3$ such that the resultant data streams is laser
noise-free. Any laser noise-free
data combination $A$ is represented by 
\begin{equation}
A = p_i V^i + q_i U^i \,,
\end{equation}
where $p_i,q_i$ are polynomials of time-delay operators
$E_i$. For the LISA configuration $L_i \sim 16.7$ seconds,
corresponding to an arm-length of 5 million km, one such known
laser noise-free data combination, expressed
in terms of 6-tuples of polynomials $\left(p_{i},\, q_{i} \right)$
is 
\begin{equation}
X = (1-E_2^2,0,E_2(E_3^2-1),1-E_3^2,E_3(E_2^2-1),0)
\end{equation}
commonly referred to as {\it Michelson} combination in the literature \cite{schilling}; in
which one arm of LISA is not used. 
In general, we have shown \cite{SNV02} that \emph{all} the data combinations
which cancel the laser frequency noise and the optical bench motion noise form
an algebraic module of {\it syzygies} over a ring of polynomials
in time delay operators $E_{i}$. This formalism
generates the noise cancellation module from the generators; linear combinations of the generators 
with polynomial coefficients in the ring generates a module. One such set of generators (which is convenient
for SNR optimization purpose) is 
$\alpha, \beta, \gamma$ and $\zeta$ (notation followed from \cite{AET99,ETA00,SNV02}). 
{\small
\begin{eqnarray}
\alpha  &=&  (1,E_{3},E_{1}E_{3},1,E_{1}E_{2},E_{2})\, ,\nonumber \\
\beta  &=&  (E_{1}E_{2},1,E_{1},E_{3},1,E_{2}E_{3})\, ,\nonumber \\
\gamma   &=&  (E_{2},E_{2}E_{3},1,E_{3}E_{1},E_{1},1)\, ,\nonumber \\
\zeta  &=&  (E_{1},E_{2},E_{3},E_{1},E_{2},E_{3})\, .\label{eq:GEN4}
\end{eqnarray}
}
We note that $\alpha, \beta, \gamma$ are cyclic permutations of each
other. The combination $\zeta$ is the symmetric Sagnac combination which
is insensitive to GW at low frequency due to its high symmetry .

\section{LISA Sensitivity Optimization}
To generate any laser noise-free data combination, in general, 4
generators are necessary but if the source is monochromatic, 
the fourth generator $\zeta$ can 
be effectively eliminated by expressing in terms of
$(\alpha,\beta,\gamma)$ as follows
\begin{equation}
(1-E_{1}E_{2}E_{3})\zeta =(E_{1}-E_{2}E_{3})\alpha +(E_{2}-E_{1}E_{3})\beta 
+(E_{3}-E_{1}E_{2})\gamma \,.
\end{equation}
except at certain frequencies which are solutions of $e^{i (L_1 + L_2
  + L_3) \Omega} = 1$.
As the maximization is possible arbitrarily 
close to the singular frequencies, the singularities do not seem to be important.

Since the difference in arm-lengths of LISA is smaller than the GW
  wavelength \cite{LISArep}, for computing the response, all the arms 
can be taken to be equal. This simplifies further analysis.
We then show that the set of 3 optimal data
  combinations having noises uncorrelated to each other can be
  obtained by linearly combining $\alpha$,$\beta$,$\gamma$. This new
  set acts as optimal when one averages over all directions and
  polarizations of the binary system. Below, we briefly summarize this
  optimization. \\
{\bf (a) Noise covariance matrix:} We define noise vectors in the Fourier domain \cite{SNV02} 
$N^{(I)}, ~I = 1, 2, 3$ for each of the generators $X^{(I)} \equiv
  \{\alpha ,\beta ,\gamma \},$ 
respectively, over the 12 dimensional complex space $\mathcal{C}^{12}$,
\begin{equation}
N^{(I)}\, \, =\, \, \left(\sqrt{S^{pf}}(2p_{i}^{(I)}+r_{i}^{(I)}),
\sqrt{S^{pf}}(2q_{i}^{(I)}+r_{i}^{(I)}),\sqrt{S^{sh}}p_{i}^{(I)},
\sqrt{S^{sh}}q_{i}^{(I)}\right)\, ,
\label{eq:NOISEVEC}
\end{equation}
 where $S^{pf}(f)=2.5\times 10^{-48}\left[f/1\: Hz\right]^{-2}\,
 Hz^{-1}$ and $S^{opt}(f) = 1.8\times 10^{-37}\left[f/1\: Hz\right]^{2}\, Hz^{-1}$ are power spectral
densities (psd) of the proof mass residual motion and the optical path noise
respectively \cite{ETA00}. The polynomials $(p_{i}^{(I)}, q_{i}^{(I)})$ corresponding to the generators $X^{(I)}$ are given in
the equation (\ref{eq:GEN4}). The $r_i^{(I)}$ polynomials are defined
through $r_1^{(I)} = -(p_1^{(I)}+E_3~q_2^{(I)}) = -(q_1^{(I)} +E_2~p_3^{(I)})$ plus cyclic
permutations for $r_2^{(I)}$ and $r_3^{(I)}$. For a given data combination, the norm of the noise vector represents
its noise psd. The noise covariance matrix
 $\mathcal{N}^{(I)}_{(J)}=N^{(I)}\cdot N^*_{(J)}$ defined for above
 generating set $X^{(I)}$  takes a symmetric form as 
\begin{equation} \label{eq:noisecov}
\mathcal{N}^{(I)}_{(J)} = \{n_d ~~{\rm for}~ I = J~{\rm and}~ n_o~~ {\rm
  for} ~I \neq J\}.
\end{equation}
{\bf (b) Signal covariance matrix:} 
The response of a GW signal for a given laser noise-free data combination $A$
is expressed in the Fourier domain \cite{SNV02} as, 
\begin{equation}
h^{(A)} (\Omega ) = \sum _{i=1}^{3}\left[p_{i}^{(A)} \left(F_{V_{i;+}}h_{+}+F_{V_{i;\times }}h_{\times }\right)
+q_{i}^{(A)} \left(F_{U_{i;+}}h_{+}+F_{U_{i;\times }}h_{\times }\right)\right] (\Omega )\, .
\label{eq:gwresp}
\end{equation}
 Here, $F_{V_{i;+/\times }}$and $F_{U_{i;+/\times }}$ are the antenna
pattern functions. 
For a binary source which may be adiabatically changing in 
frequency, the two GW amplitudes at frequency $\Omega$ are given by, 
\begin{eqnarray*}
h_{+}(\Omega ) & = & {\cal A} \left(\frac{1+\cos ^{2}\epsilon }
{2}\cos 2\psi -i\cos \epsilon \, sin2\psi \right) , \\
h_{\times }(\Omega ) & = & {\cal A} \left(-\frac{1+\cos ^{2}\epsilon }
{2}sin2\psi -i\cos \epsilon \, \cos 2\psi \right) .
\end{eqnarray*}
Here, the polarization angles $\epsilon $ and $\psi $ describe the orientation of the source and 
enter into the expressions for the polarization amplitudes. The direction of the source 
on the celestial sphere is given by the angles $\theta $ and $\phi $. 
Similar to noise covariance matrix, the signal covariance matrix
averaged over all directions and polarizations is defined as,
\begin{equation}
\mathcal{H}^{(I)}_{(J)}=\langle h^{(I)} h^*_{(J)}\rangle _{\epsilon \psi \theta \phi }\,,\label{eq:sigmat}
\end{equation}
where $\langle \; \rangle _{\epsilon \psi \theta \phi }$
represents the average over the polarizations and directions. We note
that $\mathcal{H}^{(IJ)}$ takes the same symmetric structure as the noise covariance matrix  $\mathcal{N}^{(IJ)}$ given 
in equation (\ref{eq:noisecov}). This is due to inbuilt cyclic
symmetry in this generating set $X^{(I)}$. Thus, the diagonal component of
$\mathcal{H}^{(IJ)}$ is $h_{d}$ and the off-diagonal is $h_{o}$.

Due to above symmetric structure, both $\mathcal{H}^{(IJ)}$ and
$\mathcal{N}^{(IJ)}$ can be diagonalized simultaneously. The common
eigen-observables thus obtained are given by
\begin{equation}
Y^{(1)} = \frac{1}{\sqrt{6}}\left(\alpha +\beta -2\gamma \right)\,,
\hspace{0.15in}
Y^{(2)} = \frac{1}{\sqrt{2}}\left(\beta -\alpha \right)\,, \hspace{0.15in}
Y^{(3)} = \frac{1}{\sqrt{3}}\left(\alpha +\beta +\gamma \right)\, .\label{eq:3basis}
\end{equation}
\begin{figure}
\begin{center}
\includegraphics[width=4.5in,height=2.5in]{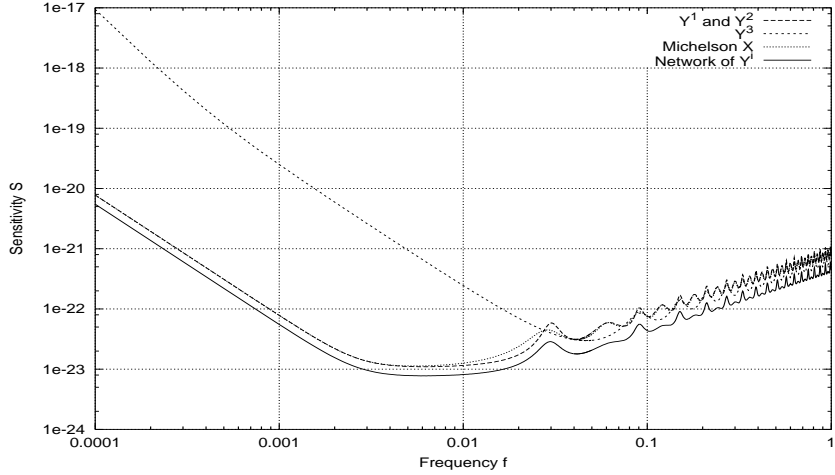}
\caption{\label{cap:Plotsens}Log Log plot of sensitivity $S$, curve as function
of $f$ after averaging over polarization and source directions for
a observation period of one year and SNR =5.}
\label{fig:sens1}
\end{center}
\end{figure}
The above eigen-observables have following properties:
\begin{itemize}
\item{Data combinations $Y^{(1)}$ and $Y^{(2)}$ have same SNR given by
$\sqrt{(h_d-h_o)/(n_d-n_o)}$. The SNR of $Y^{(3)}$ is $\sqrt{(h_d+2
    h_o)/(n_d+2n_o)}$. At any given frequency, if SNR of $Y^{(1)}$ is greater than SNR
  of $Y^{(3)}$, then $Y^{(1)}$ gives maximimum SNR whereas 
$Y^{(3)}$ gives minimum SNR or {\it vice-a-versa}. Thus, at any given
frequency, either $Y^{(1)}$ or $Y^{(3)}$ acts as optimal data
  combination amongst all the data combinations in the laser
  noise-free module.}
\item{Both $Y^{(1)}$ and $Y^{(2)}$
  perform comparable to the Michelson combination $X$ at low
  frequency. Whereas
  $Y^{(3)}$ is proportional to the symmetric Sagnac and hence
  insensitive to GW at low frequencies ($\sim$ below $3$ mHz).}
\end{itemize}
We note that averaging over directions and polarizations \cite{SNV03} results in
signal covariance matrix of rank 3. We note that if we
average over the polarizations only and obtain the optimal data 
combination for a particular direction, the signal covariance matrix 
is of rank 2. Such an optimal combination is important while
optimally tracking the source in LISA frame \cite{SNAV03}.
We may further note that in the formalism developed by Prince {\it et
  al} \cite{TTLA02}, the optimization is performed without averaging over
the source directions and polarizations, which results in signal
covariance matrix of rank 1.

\begin{figure}
\begin{center}
\includegraphics[width=4in,height=2.5in]{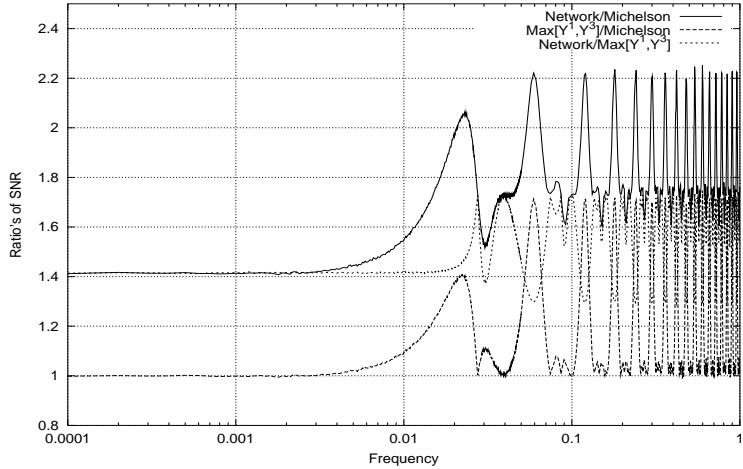}
\label{fig:net}
\caption{Plots showing the relative improvements (ratios) of SNRs for the three cases:
(i) Network SNR over the Michelson data combination (solid line).
(ii) Network SNR over the maximum of Max $[Y^{(1),}Y^{(3)}]$ (dotted line).
(iii) Max $[Y^{(1),}Y^{(3)}]$ over the Michelson (dashed line).
Here Max $[Y^{(1),}Y^{(3)}]$ is the maximum of the SNR of $Y^{(1)}$ and $Y^{(3)}$ 
over the bandwidth of LISA.}
\end{center}
\end{figure}

We have shown that either $Y^{(1)}, Y^{(2)}$ or $Y^{(3)}$ 
maximize the LISA sensitivity on an average sense and they are orthogonal {\it i.e.} they are independent random variables. 
The sensitivity of LISA can be further improved
as each of these generators can be realized as independent gravitational
wave detectors. We assume that the $Y^{(I)}$ follow the Gaussian
noise distribution and thus quadratically can combine the SNR's of
these eigen-observables to form a {\it network}-observable \cite{PDB01}. The network SNR is
given by
\begin{equation}
SNR_{network}^{2}\, \, =\, \sum_{I = 1}^{3} SNR_{(I)}^2 =\, 2
SNR_{Y(1)}^2+ SNR_{Y(3)}^2\, .
\label{eq:netSNR}
\end{equation}
The corresponding sensitivities are shown in the Fig. \ref{fig:sens1}.
In Fig. \ref{fig:net}, we have plotted the relative improvements 
in the network SNR with respect to the Michelson combination and 
the maximum of $Y^{(1)}$ and $Y^{(3)}$. At low frequencies $f \lsim 15$ mHz, the improvement of the network
SNR over the 
maximum of $Y^{(I)}$ is slightly greater than $\sqrt{2}$. This is because at low 
frequencies the data combination $Y^{(3)}$ is not very sensitive in comparison
with $Y^{(1)}$. The best improvement of factor $\sqrt{3}$ in the relative SNR is achieved at frequencies 
where all the data combinations are equally sensitive, that is, when
$SNR_{Y(1)} = SNR_{Y(3)}$.  

\section{Residual laser noise}
\begin{figure}
\begin{center}
\includegraphics[width=4.5in,height=2.3in]{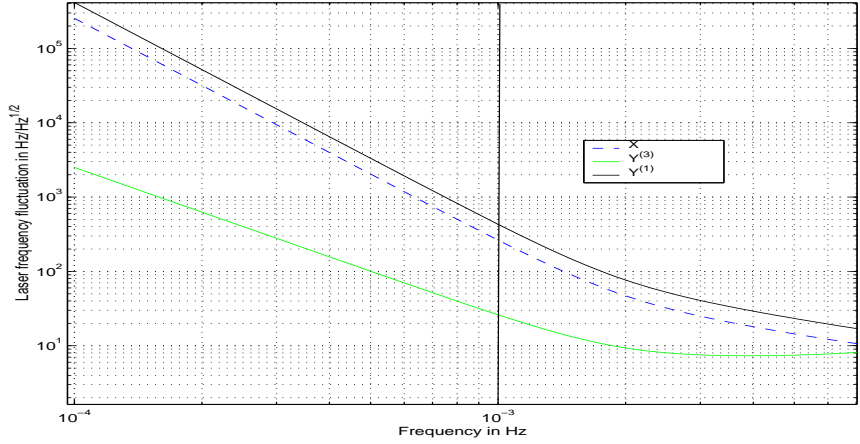}
\caption{Laser frequency stabilization $\widetilde{\Delta \nu}$ in ${\rm
    Hz/\sqrt{Hz}}$ as function of frequency for $Y^{(1)}$, $Y^{(3)}$
    and Michelson $X$ combinations, for $\Delta L = 200$ m.}
\label{fig:accur}
\end{center}
\end{figure}

As we know, it is difficult to maintain constant distance between
the three space-crafts. In general, the three arm-lengths will be different.
The actual length $L_i'$ can be estimated upto a certain accuracy.
Let the estimated arm-length be $L_i$ with an error $\Delta L_i$ such
that the actual length is $L_i' = L_i + \Delta L_i$.
Because of this unknown inaccuracy, the laser noise is not completely
canceled in a given laser noise-free data stream. 
If we demand that this residual laser phase noise level should be below the combined
noise from proof mass $S_{pf}$ and the optical path noise $S_{opt}$
then this puts an upper limit on the laser stabilization requirement
$\widetilde{\Delta \nu}$ and is given by
\begin{equation}
\widetilde{\Delta \nu} = {{\nu_0} \over{\Omega \Delta L}} \left[{{S_{pf}
    +S_{opt}} \over {\sum |p_i|^2 + |q_i|^2}}\right]^{1/2} \, .
\end{equation}
For Nd-YAG laser, $\nu_0 = 3 \times 10^{14}$ Hz and the inaccuracy
in length is assumed to be $\Delta L \sim 200$ m, the upper limit on
the laser stabilization requirement is plotted for data combinations
$Y_{(I)}$ in Fig. {\ref{fig:accur}}. For this assumed inaccuracy in arms,
the $Y^{(3)}$ combination demands that the laser frequency stabilization 
be at least as good as ${\widetilde{\Delta \nu}} \sim 25 ~{\rm
  Hz}/\sqrt{\rm Hz}$ at $1$ mHz. While for $Y^{(1)}$, the requirement
is much less stringent on frequency stabilization. The laser stabilization
requirement scales linearly with the assumed arm-length inaccuracy.
 

\section{Conclusion}
For any frequency in LISA band, we show that the optimal laser noise-free data
combinations (when averaged over all directions and polarizations of
GW source) are nothing but the eigen-data combinations of
noise-covariance matrix. Since these combinations have
uncorrelated noise, their SNR's can be combined quadratically to
improve LISA sensitivity. The improvement varies from $40 \%$ to
$100 \%$ with respect to Michelson data combination. We further show
that with our demand that the residual laser noise should be less than the
proof mass noise and optical path noise, the laser frequency stabilization
requirement varies inversely proportional to the arm-length inaccuracy.
The stringent demand on the laser stabilization requirement is
${\widetilde{\Delta \nu}} \sim 25~{\rm Hz}/\sqrt{\rm Hz}$ at $1$ mHz
for inaccuracies in arm-length of $200$ m. 

\section*{Acknowledgments}
KRN,SVD,JYV would like to thank IFCPAR under which this work has been
carried out. AP would like to thank CNRS, Observatoire de la C\^ote d'Azur
for Henri Poincare Fellowship grant
and european union for providing the local hospitality at Moriond.

\end{document}